\documentclass[aps, prd, amsmath, amssymb, a4paper, superscriptaddress, preprintnumbers, nofootinbib, showpacs, 11pt]{revtex4-1}

\usepackage{amsmath}
\usepackage{amsfonts}
\usepackage{amsthm}
\usepackage{amssymb}
\usepackage{bbm}
\usepackage{graphicx}
\usepackage{hyperref}

	\newcommand{\beq}{\begin{equation}}
	\newcommand{\eeq}{\end{equation}}

	\newcommand{\e}{\mathrm{e}}

\begin{document}

	\title{UV dimensional reduction to two from group valued momenta}
	\date{\today}
	
	\author{Michele Arzano}
	\email{michele.arzano@roma1.infn.it}
	\affiliation{Dipartimento di Fisica, Universit\`a di Roma La Sapienza,\\ P.le Aldo Moro 5, I-00185 Rome, Italy}
	\author{Francisco Nettel}
	\email{francisco.nettel@roma1.infn.it, fnettel@nucleares.unam.mx}
	\affiliation{Dipartimento di Fisica, Universit\`a di Roma La Sapienza,\\ P.le Aldo Moro 5, I-00185 Rome, Italy}
	\affiliation{Instituto de Ciencias Nucleares, Universidad Nacional Aut\'onoma de M\'exico,
AP 70543, M\'exico, CDMX 04510, Mexico}

\begin{abstract}
We describe a new model of deformed relativistic kinematics based on the group manifold $U(1) \times SU(2)$ as a four-momentum space. We discuss the action of the Lorentz group on such space and and illustrate the deformed composition law for the group-valued momenta. Due to the geometric structure of the group, the deformed kinematics is governed by {\it two} energy scales $\lambda$ and $\kappa$. A relevant feature of the model is that it exhibits a running spectral dimension $d_s$ with the characteristic short distance reduction to $d_s =2$ found in most quantum gravity scenarios.

\end{abstract}


\maketitle

\section{Introduction}  
\label{sec:intro}

Is space-time fundamentally four-dimensional or there exist regimes in which its effective dimensionality can change? In the past few years hints in favour of the second scenario have been accumulating from a variety of approaches to quantum gravity \cite{Ambjorn:2005db,Horava:2009if,Carlip:2009kf,Modesto:2008jz,Benedetti:2008gu,Benedetti:2009ge,Sotiriou:2011aa,Calcagni:2013vsa,Calcagni:2010bj,Arzano:2013rka,Alesci:2012,Arzano:2014jfa,Carlip:2015mra}. These works have shown evidence for a {\it reduction} of the number of dimensions approaching the Planck scale and, strikingly, most of them agree on the prediction of an effectively two-dimensional (quantum) spacetime in the UV. A tool widely used in all these analyses is the spectral dimension, a quantity which captures the dimensionality of (Euclidean) quantum spacetime as probed by a fictitious diffusion process. 

A running spectral dimension at short distances can be modelled via the introduction of UV corrections to the usual quadratic Laplacian which, in momentum space, reflect a modification of the energy-momentum dispersion relation \cite{Sotiriou:2011aa}. Such modified dispersion relations are in general associated to a violation of Lorentz invariance like, for example, in Horava-Lifschitz models \cite{Horava:2009if} and their close relatives \cite{Sotiriou:2010wn}. This is not the only possibility however. Deformed dispersion relations also appear naturally in theories with ``deformed" relativistic kinematics at the Planck scale. 
These models were first introduced as a way to modify the algebraic structure of relativistic symmetries to introduce a fundamental, observer independent, Planckian energy scale \cite{AmelinoCamelia:2000mn}, and were later suggested as effectively describing the imprint of the ``flat spacetime limit" of quantum gravity on relativistic kinematics \cite{AmelinoCamelia:2003xp,KowalskiGlikman:2004qa}. Their basic features, a non-abelian composition law for momenta and a non-linear action of boosts saturating at the Planck scale, were inspired by existing mathematical deformations of the Poincar\'e algebra based on the technology of quantum groups and Hopf algebras \cite{Lukierski:1991,Lukierski:1992,Lukierski:1994}. It was soon realized that these kinematical properties can be obtained \cite{Majid:1994,Nowak:2002,Kowalski:2002,Kowalski:2003} by replacing the usual Minkowski energy-momentum space with a {\it non-abelian Lie group} equipped with a natural action of the Lorentz group.  For models based on the so-called $\kappa$-Poincar\'e algebra, the momentum space was identified with the Lie group $AN(3)$, a subgroup of the five dimensional Lorentz group, $SO(4,1)$. The $AN(3)$ group, as a manifold, is ``half" of four-dimensional de Sitter space whose cosmological constant determines a fundamental Planckian energy scale $\kappa$ associated to the deformation. On this space there exists a natural action of the four-dimensional Lorentz group thanks to the Iwasawa decomposition of the five-dimensional Lorentz group $SO(4,1)\sim SO(3,1) AN(3)$ \cite{KowalskiGlikman:2004tz}. 

Diffusion processes defined on the Euclidean version of the $AN(3)$ group manifold momentum have been studied in detail and generally exhibit a running of the spectral dimension in the UV \cite{Arzano:2014jfa}. Such change in dimensionality can lead to various short-distance limits which differ from the IR value of 4 and can, in principle, be larger or smaller than 4 depending on the specific choice of Laplacian. In particular for what concerns the UV value of 2, suggested by most quantum gravity scenarios, it has been showed that it can be reached only for a fine tuned choice of Laplacian \cite{Amelino-Camelia:2013cfa,Amelino-Camelia:2013gna}. This is certainly a drawback if such models are to be understood as possible scenarios for Planck scale kinematics, effectively modelling quantum gravity effects.

In this letter we introduce a new framework for deformed relativistic kinematics based on a Lie group momentum space which, among others, has the desirable property of naturally leading to a dimensional reduction to 2 in the UV, without requiring an ad hoc choice of the Laplacian. Our model is based on the four-dimensional momentum Lie group $U(1) \times SU(2)$ which, as a manifold, is the cartesian product of a circle times the three-sphere: $S^1\times S^3$. Motivation for considering such momentum space comes from the study of particles coupled to Einstein gravity in three spacetime dimensions. Early analyses on the kinematics of such particles led 't Hooft, in the mid 90s, to consider quantization on a $S^1\times S^2$ energy-momentum space \cite{'tHooft:1996uc} (see also \cite{Welling:1997qz}). Various following studies, motivated also by the formulation of three-dimensional gravity as a Chern-Simons theory \cite{Carlip:1989nz,Bais:1998yn,Bais:2002ye}, established that the energy-momentum space of gravitating point particles should be taken to be the Lie group $SL(2, \mathbb{R})$ which, as a manifold, is the three-dimensional anti-de Sitter space \cite{Matschull:1997du}. Such group valued momenta can be obtained by exponentiating ordinary three-momenta using Newton's constant $G$ to obtain a dimensionless argument in the exponential. This is achieved via the vector-space isomorphism $\mathbbm{R}^{1,2}\simeq \mathfrak{sl}(2,\mathbbm{R})$, between three dimensional Minkowski space and the space of real, traceless, two by two matrices, the vector space of the Lie algebra $\mathfrak{sl}(2,\mathbbm{R})$. An ordinary particle's three-momentum can indeed be represented by a matrix
\beq\label{3mom}
p =
\begin{pmatrix}
           p_2   & p_1+ p_0 \\
      p_1-p_0 &  -p_2 
\end{pmatrix} 
\eeq
on which Lorentz transformations $g\in SL(2, \mathbb{R})$ act via the adjoint action $\mathnormal{Ad}_g\,(p) = g\, p\, g^{-1}$ and whose determinant provides the invariant mass-shell condition $\mathrm{det}(p) = (p_0)^2 - (p_1)^2 - (p_2)^2 = m^2$. According to \cite{Bais:2002ye,Matschull:1997du} when the particle is coupled to gravity its momentum will be given by an element of $SL(2, \mathbb{R})$ obtained by exponentiating \eqref{3mom}
\beq\label{3gmom}
h = e^{4\pi G p}\,.
\eeq
The adjoint action of the Lorentz group on $\mathfrak{sl}(2,\mathbbm{R})$ gets mapped to the action by conjugation of $SL(2, \mathbb{R})$ on itself \cite{Baez:2006un} and the mass-shell invariant is determined by the trace of the group element \eqref{3gmom}. These are the basic ingredients of the deformation of three-dimensional relativistic kinematics associated with the $SL(2, \mathbb{R})$ momentum group manifold \cite{Arzano:2014ppa}.

The model we propose here, based on the momentum group manifold $U(1) \times SU(2)$, can be seen as a direct four-dimensional generalization of the three-dimensional example above. As in the three-dimensional case, ordinary four-momenta can be described by complex two by two matrices via the vector space isomorphism between $\mathbbm{R}^{1,3}$ and the Lie algebra $\mathfrak{u}(1) \oplus \mathfrak{su}(2)$. On these matrix four-momenta one has a natural action of the Lorentz group and the determinant again provides a Lorentz invariant object which one identifies with the mass-shell condition. Introducing two fundamental energy scales one can exponentiate such Lie algebra elements to obtain group valued momenta. This is the starting point of our work.

We begin, in the next Section, by describing the basics of the Lie group $U(1) \times SU(2)$ as momentum space introducing two useful parametrizations of its elements. As a next step we illustrate the action of the Lorentz group, its explicit form for a given group parametrization, and derive the non-abelian composition of momenta related to the group multiplication. We finally calculate the spectral dimension associated with the $U(1) \times SU(2)$ momentum space and show that at small diffusion times it flows to 2, a feature found in most approaches to the description of spacetime at the Planck scale.

\section{Introducing the $U(1) \times SU(2)$ momentum space}  
\label{sec: U(1) SU(2) momentum space}

We begin with a brief description of the $U(1) \times SU(2)$ Lie group as a momentum space starting from the associated Lie algebra. 
A basis of the Lie algebra $\mathfrak{u}(1) \oplus \mathfrak{su}(2)$ is given by the four matrices $X^\mu = \{i \mathbbm{1}, i \sigma^1, -i \sigma^2, i \sigma^3\}$, where $\sigma^i$, $i=1,2,3$, are the Pauli matrices
	\beq \label{paulimat}
	\sigma^1 = 
	\begin{pmatrix} 
	0 && 1 \\
	1 && 0 
	\end{pmatrix}, \qquad
	\sigma^2 = 
	\begin{pmatrix}
	0 && - i \\
	i && 0 
	\end{pmatrix}, \qquad
	\sigma^3 = 
	\begin{pmatrix}
	1 && 0 \\
	0 && -1
	\end{pmatrix}\,,
	\eeq
and $\mathbbm{1}$ is the identity matrix. The Lie brackets for $\mathfrak{u}(1) \oplus \mathfrak{su}(2)$ are given by
	\beq  \label{u2brackets}
	[X^0, X^i] = 0, \qquad [X^i, X^j] = 2 \epsilon^{ijk} X^k,
	\eeq
with $\epsilon^{123} = 1$ and the matrix form for an element $K = k_\mu X^\mu\in \mathfrak{u}(1) \oplus \mathfrak{su}(2)$ is given by
	\beq \label{Pelement}
	K = i
	\begin{pmatrix}
	k_0 + k_3 && k_1 + i k_2 \\
	k_1 - i k_2 && k_0 - k_3
	\end{pmatrix},
	\eeq
A group element $h \in U(1) \times SU(2)$ can be obtained by exponentiating the matrix $K$. Such group element can be written as $h = h_1 h_2$, with $h_1 \in U(1)$ and $h_2 \in SU(2)$ and the exponential map provides a global parametrization of the group given by {\it exponential coordinates} $k_{\mu}$ which express $h_1 \in U(1)$ and $h_2 \in SU(2)$, respectively, as
	\beq \label{expcoord}
	h_1 = \exp \left(\frac{k_0}{\lambda} X^0\right) \qquad \text{and} \qquad h_2 = \exp \left(\frac{k_i}{\kappa}X^i\right)\, .
	\eeq
Notice that we introduced two energy scales $\lambda$ and $\kappa$, in order to have coordinates on the group manifold with the dimension of energy. The explicit matrix form of the group element in these coordinates is given by
	\beq
	h = e^{i \frac{k_0}{\lambda}}
	\begin{pmatrix}
	\cos \frac{|k|}{\kappa} + i\, \frac{k_3}{|k|}\, \sin \frac{|k|}{\kappa}  &&  \frac{i}{|k|}\, \sin \frac{|k|}{\kappa}\, (k_1 + i k_2) \\
	 \frac{i}{|k|}\,\sin \frac{|k|}{\kappa}\, (k_1 - i k_2) && \cos \frac{|k|}{\kappa} - i\, \frac{k_3}{|k|}\, \sin \frac{|k|}{\kappa} 
	\end{pmatrix},
	\eeq
where $|k| = \sqrt{k_1{}^2 + k_2{}^2 + k_3{}^2}$. As in other models of curved four-momentum space, different sets of coordinates on the group manifold correspond to different choices of momenta and in general lead to different kinematics.
For example, we can define new momenta $p_{\mu}$ associated to {\it embedding coordinates} by introducing the following parametrization of the group elements
	\beq \label{h1}
	h_1 = v \mathbbm{1}+ \frac{p_0}{\lambda} X_0 \qquad \text{and} \qquad h_2 = u\mathbbm{1} + \frac{p_i}{\kappa} X^i\,.
	\eeq
The real parameters $v$ and $u$ are not independent from the momentum variables $p_0$ and $p_i$, respectively, but are subject to unimodular conditions which ensure that $h_1 \in U(1)$ and $h_2 \in SU(2)$ leading to the following relations
	\beq \label{embeddcond0}
	(v \lambda)^2 + p_0{}^2 = \lambda^2 \qquad (u \kappa)^2 + |p|^2 = \kappa^2\,.
	\eeq
Notice that these coordinates are not global since different sign choices for $v$ and $u$ cover different patches of the manifold. For instance, choosing the positive sign for both $v$ and $u$ 
	\beq \label{embeddcond}
	v = \sqrt{1 - \frac{p_0^2}{\lambda^2}} \qquad \text{and} \qquad u = \sqrt{1 - \frac{|p|^2}{\kappa^2}},
	\eeq
where $|p|^2 = p_1{}^2 + p_2{}^2 + p_3{}^2$, we cover the ``upper half'' of $S^1$ and $S^3$, respectively. The equations in  \eqref{embeddcond0} describe the manifolds of the two groups $U(1)$ and $SU(2)$ within the embedding space, respectively a one-sphere and a three-sphere. Notice that the two scales $\lambda$ and $\kappa$ determine the radii of the manifolds $S^1$ and $S^3$. 

\section{The action of Lorentz group} 
\label{sec:Lorentz_action}

In order for the coordinates on the group $U(1) \times SU(2)$ to have a meaning as actual energies and momenta we have to specify their relativistic transformations i.e. to define an action of the Lorentz group on $U(1) \times SU(2)$. To do so let us recall that {\it ordinary} relativistic transformations can be described using a matrix representation of the four-momentum space $\mathbb{R}^{1,3}$ via the isomorphism of vector spaces with the Lie algebra $\mathfrak{u}(1) \oplus \mathfrak{su}(2)$ given by \eqref{Pelement}. Such action is based on the spinorial representation of the Lorentz group generated by the following basis for $sl(2, \mathbb{C})$, $\{j_i, k_i\} = \{\frac{i}{2} \sigma^1, - \frac{i}{2} \sigma^2, \frac{i}{2} \sigma^3, \frac{1}{2} \sigma^1, - \frac{1}{2} \sigma^2, \frac{1}{2} \sigma^3 \}$. Notice that the determinant of the four-momentum matrix $K = k_\mu X^\mu\in \mathfrak{u}(1) \oplus \mathfrak{su}(2)$ is $\mathrm{det}\,(K) = - k_0^2 +  \sum k_i^2$. The action of $g \in SL(2,\mathbb{C})$ on $K$ is defined by
	\beq \label{lorentzonP}
	K \longrightarrow K' = g K g^*,
	\eeq
where $g^*$ is the conjugate transpose of $g$. The determinant of $K$ is obviously invariant under the action of the group element $g$ and indeed $\det (K) = - k_0^2 +  \sum k_i^2$ reproduces the familiar mass-shell condition if one identifies $\det (K)$ with the (minus) mass squared. This description of the action of the Lorentz group on Minkowski space, suggests a natural definition of Lorentz transformations on the group manifold $U(1) \times SU(2)$ through 
	\beq \label{lorentzonh}
	h = \exp(K) \longrightarrow h' = \exp(gKg^*) = \exp(K').
	\eeq
Choosing a set of coordinates on the group manifold one can write down an explicit form of such transformation. For example, choosing four-momenta defined by exponential coordinates, a Lorentz boost in the 1-direction is given by ordinary expression
	\begin{align} \label{boostk1}
	k'_0 &= k_0 \cosh \beta + k_1 \sinh \beta, \nonumber \\
	k'_1 &= k_1 \cosh \beta + k_0 \sinh \beta, \nonumber \\
	k'_i &= k_i \qquad i = 2,3.
	\end{align}
The same applies to the mass-shell relation which has the usual form
	\beq \label{DRexp}
	-k_0{}^2 +  |k|{}^2 + m^2 = 0\,,
	\eeq
reflecting an undeformed mass Casimir invariant associated to the translation generators of the associated Poincar\'e group.	

Notice that choosing a different parametrization of the group, the action of boosts will be in general non-linear. For momenta defined by the embedding coordinates $p_{\mu}$ the counterpart of the action \eqref{boostk1}, for example, can be easily derived using the following relations between cartesian and embedding coordinates
	\beq \label{exptocart0}
	p_0 = \lambda \sin \frac{k_0}{\lambda}\,, \qquad  v = \cos \frac{k_0}{\lambda},
	\eeq
and 
	\beq  \label{exptocart1}
	p_i = \kappa \frac{\sin \frac{|k|}{\kappa}}{|k|}\ k_i \,,\qquad  u = \cos \frac{|k|}{\kappa}\,.
	\eeq
Let us point out that exponential momenta $k_{\mu}$ are subject to ordinary relativistic transformations and thus, via rotations and boost, can assume {\it any} real values. 
In particular energies can be boosted above the ``Planckian" value of $2\pi \lambda$ and the natural question that arises is whether we should identify boosted energies mod $2\pi \lambda$. As it can be easily verified, for on-shell momenta such identification of energies would be unphysical since it would force the new boosted momentum to a another shell with a {\it different mass}.\footnote{To see this let us boost along the $x$-axis a four momentum at rest $k_{\mu}=(m,0,0,0)$ to a new four-momentum with energy $2\pi \lambda + \epsilon$ and make the identification $2\pi \lambda + \epsilon = \epsilon = k'_0$. The new four-momentum will now have a linear component along the $x$-direction: $k'_1 = \sqrt{-m^2+(2\pi \lambda + \epsilon)^2}$ and thus $(k'_0)^2-(k'_1)^2\neq  m^2$.} Thus if exponential momenta $k_{\mu}$ are to be identified with momenta of physical particles their values under boosts can be arbitrary i.e. the domain of exponential momenta, seen as complex logarithms, is a Riemann surface rather than a principal branch. This, in principle, is not the case for other choices of coordinates where the non-linear action of Lorentz transformations can be such that under an infinite boost energy and/or momentum reach maximum values set by the deformation scales. This can be immediately verified for momenta defined by cartesian coordinates, where, for example, energies are bounded by the value of $\lambda$. Such behaviour is characteristic of models known as ``doubly special relativity" \cite{AmelinoCamelia:2000mn,Magueijo:2001cr}. We postpone a detailed discussion of such ``deformed" non-linear boost transformation and their properties to an upcoming work. For the moment we should just stress that, as it can be easily checked {\it at first order} in $1/\kappa$ and $1/\lambda$, the Lorentz transformations and mass-shell relation for exponential and cartesian momenta, $k_{\mu}$ and $p_{\mu}$, coincide and are just the familiar relativistic ones.

\section{Composition of momenta}
\label{sec:compomomenta}

A peculiar feature of models of deformed kinematics based on a momentum group manifold is the non-abelian composition of four-momenta inherited from the group multiplication law. As argued in \cite{Arzano:2016yir} such composition law reflects the way momenta, seen as quantum numbers, add for multiparticle states. As for Lorentz transformations and mass Casimir, the explicit form of the composition law depends on the choice of coordinates for the momentum manifold. Denoting by $p_{\mu}(h)$ the four-momentum coordinates associated to the group element $h$, the non-abelian composition law is defined by
\beq
p_{\mu}(h_1) \oplus p_{\mu}(h_2) \equiv p_{\mu}(h_1 h_2)\,.
\eeq 
Let us write down the explicit form of such composition law for cartesian and exponential coordinates. In cartesian coordinates, using the following notation for  momenta associated to different group elements, $p^1_\mu=p_\mu(h_1)$ and $p^2_\mu=p_\mu(h_2)$, the composition law is given by
	\begin{align} \label{compcart}
	p^1_0 \oplus p^2_0 &= p^1_0\, \sqrt{1 - \left(\frac{p^2_0}{\lambda}\right)^2}+ p^2_0\, \sqrt{1 - \left(\frac{p^1_0}{\lambda}\right)^2}, \nonumber \\
	p^1_i \oplus p^2_i &= p^1_i\ \sqrt{1 - \left(\frac{|p^2|}{\kappa}\right)^2}  + p^2_i\ \sqrt{1 -  \left(\frac{|p^1|}{\kappa}\right)^2}  + \frac{1}{\kappa} \sum_{j,k} \epsilon_{ijk}\ p^1_j p^2_k\, .
	\end{align}
Notice that, since the $U(1)$ energy component of the group commutes with the $SU(2)$ spatial part, the composition rule above is ``decoupled" in energy and momentum unlike the composition of four-momenta found in models of $\kappa$-deformed kinematics in which, for example, the non-abelian addition of spatial momenta involves an energy-dependent factor \cite{Arzano:2010jw}. In exponential coordinates, with $k^1_\mu=k_\mu(h_1)$ and $k^2_\mu=k_\mu(h_2)$, the composition law for the energy-component of momentum remains undeformed while the one for spatial momenta becomes quite involved:
	\begin{align} \label{compexp}
	k^1_0 \oplus k^2_0 &= k^1_0 + k^2_0, \nonumber \\
	k^1_i \oplus k^2_i &= \frac{1}{f(k^1,k^2)} \left[k^1_i\ \frac{\sin \frac{|k^1|}{\kappa}}{|k^1|} \cos \frac{|k^2|}{\kappa}  + k^2_i\ \frac{\sin \frac{|k^2|}{\kappa}}{|k^2|} \cos \frac{|k^1|}{\kappa}  + \frac{\sin \frac{|k^1|}{\kappa}}{|k^1|} \frac{ \sin \frac{|k^2|}{\kappa}}{|k^2|} \sum_{j,k} \epsilon_{ijk}\ k^1_j k^2_k \right],
	\end{align}
where 
	\beq 
	f(k^1,k^2) = \frac{ \left[ 1 - \left( \cos \frac{|k^1|}{\kappa} \cos \frac{|k^2|}{\kappa} - \frac{ \sin \frac{|k^1|}{\kappa}}{|k^1|} - \frac{ \sin \frac{|k^2|}{\kappa}}{|k^2|} \sum_i k^1_i k^2_i \right)^2 \right]^{1/2}}{\kappa \arccos \left( \cos \frac{|k^1|}{\kappa} \cos \frac{|k^2|}{\kappa} - \frac{ \sin \frac{|k^1|}{\kappa}}{|k^1|}  \frac{ \sin \frac{|k^2|}{\kappa}}{|k^2|} \sum \limits_i k^1_i k^2_i \right)}.
	\eeq
Again an important question to address is whether we should consider the sum of energies in \eqref{compexp} to be defined mod $2\pi\lambda$. In the previous Section we have seen that the restriction of exponential momenta to a branch of the logarithm is unphysical since, for example, a periodic identification of energies would lead to a jump to another mass-shell at each branch point. Thus at the kinematical level one should think of the composition of energies in \eqref{compexp}  as an addition of complex logarithms defined on a Riemann surface and thus {\it not} to be defined mod $2\pi \lambda$.

Finally let us notice that the deformed addition rules above are such that in the ``flat" limit $\lambda, \kappa \to \infty$ reduce to the usual abelian addition of momenta. It is instructive to write down the composition law at first order in $1/\lambda$ and $1/\kappa$ which turns out to have the same form for both group parametrization and reads
      \begin{align}  \label{complaw1order}
	p^1_0 \oplus p^2_0 &= p^1_0 + p^2_0, \nonumber \\
	p^1_i \oplus p^2_i &= p^1_i + p^2_i + \frac{1}{\kappa} \epsilon_{ijk}\, p^1_j p^2_k.
	\end{align}
As discussed in the previous section, at first order in $1/\lambda$ and $1/\kappa$ the action of boosts and the mass-shell condition coincide for exponential and cartesian coordinates. Thus the composition law \eqref{complaw1order} together with ordinary mass shell and undeformed boost provide a minimal model of deformed kinematics in which only the composition law of spatial momenta is affected. 

Let us mention that at a mathematical level the non-abelian composition of momenta reflects a non trivial action of translation generators on tensor product representations (or multiparticle states) encoded in a non-trivial {\it coproduct} (see e.g. \cite{Arzano:2014cya} for a detailed discussion). To complete the picture one should also specify the action of Lorentz transformation on such states or, equivalently, describe how the sum of group valued momenta transforms under boost.
The action \eqref{lorentzonh} can be used to infer the action of the Lorentz group on the sum of group-valued momenta which turns out to be undeformed. In fact it is immediate to see that for a generic Lorentz transformation $(h_1h_2)' \equiv h'_1 h'_2$, i.e. boosting the sum of two momenta $h_1$ and $h_2$ is equivalent to taking the sum of the boosted individual momenta. This indicates that, in mathematical language, the {\it co-algebra sector} of the Lorentz algebra, in this context, is trivial. This is in analogy with models of three-dimensional deformed kinematics based on the $SL(2, \mathbb{R})$ momentum space, whose Hopf algebraic sector is described by the {\it quantum double} of $SL(2, \mathbb{R})$ \cite{Bais:1998yn,Bais:1998yn,Arzano:2014ppa}.

\section{Dimensional reduction}
\label{sec:spectral}

In the Introduction, we pointed out that models of deformed kinematics based on a non-trivial geometry of momentum space, generally feature a running spectral dimension $d_s$ in the UV. 
The specific behaviour of the running dimensionality at short distances is dictated by the geometry of momentum space, via the integration measure, and, as in models with Lorentz invariance violation, by the choice of Laplacian and the associated mass Casimir invariant. Here we calculate the spectral dimension associated to the $U(1)\times SU(2)$ momentum space for the most natural choice of Laplacian, the one associated with the mass-shell relation \eqref{DRexp}. As we will see the running spectral dimension will exhibit the characteristic UV-limit reduction to $d_s =2$ found in most quantum gravity scenarios. 

Let us first recall how the spectral dimension is defined as the effective dimension probed by a diffusion process defined on a manifold. The diffusion process can be described by the probability density $\rho(x,x_0;s)$ of reaching a point $x$ from a starting point $x_0$ in a diffusion time $s$ with the initial condition $\rho(x,x_0;s) = \delta(x - x_0)$. The return probability $P(s)$ is obtained averaging over all the points in the manifold the probability density of diffusion at $x = x_0$
	\beq \label{PofSdef}
	P(s) = \frac{1}{V} \int d^d x\, \rho(x,x;s)\,,
	\eeq
To  $P(s)$ one can associate the spectral dimension defined as
	\beq \label{specdim}
	d_s = -2 \frac{d \ln P(s)}{d \ln s}.
	\eeq 
The return probability can be explicitly calculated by solving the {\it heat equation} $\partial \rho/ \partial s = \Delta \rho$ for the probability density $\rho$, where $\Delta$ is the Laplacian associated to the Euclidean version of the manifold under consideration. A general solution of the diffusion equation can be written as
	\beq \label{probdens}
	\rho(x, x_0;s) = \int \frac{dp_0\, d^{d-1}p}{(2\pi)^d} \, \e^{ip(x - x_0)} \e^{-sC(p)},
	\eeq
where $C(p)$ is the momentum space representation of the Laplacian which for Minkowski space reproduces the Wick-rotated mass-shell condition associated to the mass Casimir of the Poincar\'e algebra. We can thus write the return probability $P(s)$ as an integral over momentum space of a simple function of the momentum space Laplacian $C(p)$ or energy-momentum dispersion relation
	\beq  \label{Pofs}
	P(s) = \int \frac{dp_0\, d^{d-1}p}{(2 \pi)^{d}} \e^{-s C(p)}.
	\eeq
As discussed in detail in \cite{Sotiriou:2011aa}, UV-modifications of the energy-momentum dispersion relation, reflecting in a deformed momentum space Laplacian $C(p)$, are in general associated with a running spectral dimension at short distances. In models with broken Lorentz invariance, like e.g. Horava-Lifschitz gravity, the UV-modified form of $C(p)$ essentially captures all the non-trivial features of the short distance behaviour of the spectral dimension. The deformed kinematics associated to the group manifold structure of momentum space introduces an additional feature in \eqref{Pofs}: the possibility of a non-trivial integration measure in energy-momentum space. Indeed, when the latter is given by a Lie group the integration must be performed using the Haar measure. 
Such measure can be easily written starting from the Lebesgue measure on the emebdding six-dimensional Minkowski space multiplied by a delta function imposing the manifold constraint. For $U(1) \times SU(2)$ the Haar measure is simply given by the product of the Haar measures on the two groups and, in cartesian coordinates, is given by
	\beq \label{haarcart}
	d\mu_h = \frac{1}{\lambda \kappa^3} \frac{dp_0}{\sqrt{1 - \frac{p_0{}^2}{\lambda^2}}} \frac{d^3p}{\sqrt{1 - \frac{|p|^2}{\kappa^2}}}.
	\eeq
We calculate here the spectral dimension associated to the kinematics described by {\it exponential coordinates}, for which momenta can assume any real value (ordinary boosts) and the energy momentum dispersion relation \eqref{DRexp} is undeformed. 
The Haar measure for such four-momenta is given by
	\beq \label{haarexp}
	d\mu_h = \frac{1}{\lambda \kappa} \frac{\sin^2 \frac{|k|}{\kappa}}{|k|^2}\, dk_0\, d^3k\,,
	\eeq
and the return probability thus reads 
	\beq \label{Pofs2}
	P(s) = \int \frac{dk_0 d^3k}{(2\pi)^4\, \lambda\kappa}\ \frac{\sin^2 \frac{|k|}{\kappa}}{|k|^2}\ \e^{-s(k_0{}^2 + |k|^2)}\,,
	\eeq
where we used the Wick-rotated dispersion relation $C(k) = k_0{}^2 + |k|^2$.  The integral can be easily calculated using polar coordinates for the spatial part
	\beq \label{PofsSphe}
	P(s) =  \int  \frac{dk_0 dk_r}{(2\pi)^3}\frac{2}{\lambda\kappa}\ \sin^2 \frac{k_r}{\kappa}\ \e^{-s(k_0{}^2 + k_r{}^2)},
	\eeq 
\begin{figure}[htb]
\centering
\includegraphics[scale=1]{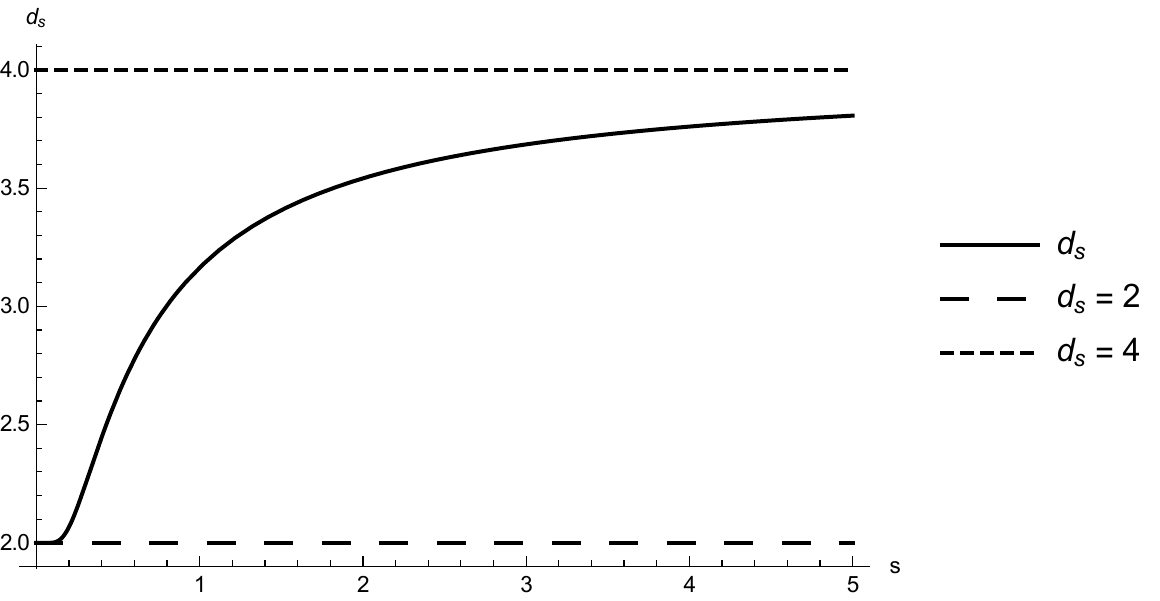}\caption{Spectral dimension for the momentum space $U(1) \times SU(2)$ ($\lambda = \kappa = 1$).}
\label{fig1}
\end{figure}	
where $k_r$ is the radial component and the momentum components $k_0, k_r \in [0,\infty)$. Fixing $\lambda = \kappa = 1$ we obtain after integration
	\beq \label{PofsRes}
	P(s) = \frac{1 - \e^{-1/s}}{32 \pi^2 s}\,,
	\eeq
and, from the definition \eqref{specdim}, we can write the spectral dimension $d_s$ as a function of diffusion time $s$ as
  \beq \label{specdimu1su2}
	d_s = 2 + \frac{2}{s(\e^{1/s} - 1)}.
	\eeq
In figure \ref{fig1} we plotted the running spectral dimension. Its long and short-distance limits can be easily calculated. The infrared limit is $\lim_{s \to \infty} d_s = 4$, as one would expect, while for the short distance limit gives $\lim_{s \to 0} d_s = 2$, which is the same two-dimensional behaviour found in various quantum gravity scenarios.\\ 

An important point to be stressed is that the Haar measure defined on our four-momentum space transforms non-trivially under Lorentz boosts. Such lack of Lorentz invariance of the momentum space measure could be seen as a drawback of the model. However it should be stressed that subtle deviations from Lorentz invariance are a characteristic feature of deformed kinematics based on curved momentum space see e.g. \cite{Amelino-Camelia:2013cfa, Arzano:2014jua}. These arise typically from a tension between the action of Lorentz transformation and the domain over which the momentum coordinates are defined. In our model the structure of momentum space leads to a natural integration measure which is invariant only under the restricted set of symmetries encoded by the momentum group manifold. We expect that a generalization of the model presented here to include spin degrees of freedom in the phase space would require the adoption of a larger group as momentum space and thus could potentially offer a way to define a Lorentz invariant measure on the deformed phase space. We will address this point in future work.

\section{Summary}
We introduced the Lie group $U(1) \times SU(2)$ as a energy-momentum space and laid down the basics of a new model for deformed kinematics in four spacetime dimensions. The main motivation to consider such Lie group as a four-momentum space comes from the analogy with the description of gravitating point particles in three spacetime dimension. In this context various studies have shown that group valued momenta replace ordinary three-momentum vectors in the description of the particles phase space.
Such group valued momenta can be seen as the {\it exponentiated} version of ordinary three-momenta represented by two by two, real, traceless matrices. Likewise $U(1) \times SU(2)$-valued momenta can be obtained by exponentiating ordinary four-momenta written in terms of two-by-two hermitian matrices. While in the three dimensional case Newton's constant provides the mass-scale which governs the geometry of the group in our case we have two, in principle different, fundamental energy scales associated with the radii of the two groups $U(1)$ and $SU(2)$.

A remarkable feature of the model we described is that it naturally reproduces a spectral dimensiona running to 2 at short distances, matching the results found in various quantum gravity scenarios. Another desirable property of the $U(1) \times SU(2)$ momentum space is that it provides a natural splitting between energy and momentum space reflecting the possibility of a non-abelian composition of four-momenta which {\it does not} mix spatial momenta and energy as in models of $\kappa$-deformed kinematics. At the field theoretic level this natural separation between energy and momentum will also reflect in a lack of ambiguity in the definition of plane waves. The need to specify a normal ordering between space and time part of plane waves has indeed been one of the technical difficulties which limited the development of $\kappa$-deformed field theories.

One of the important points we leave for future investigations is the determination of the Hopf algebraic structures which encodes the non-trivial picture of the Poincar\'e group obtained from adopting $U(1) \times SU(2)$ as a relativistic momentum space. On the more physical side it will be interesting to determine whether for some specific parameterizations of the momentum group manifold one can obtain non-linear actions of boosts which lead to maximal values of energy and/or spatial momenta set by the two deformation scales, as is the case of other models of deformed kinematics based on group valued momenta. We should also point out that, as showed in a series of works \cite{Amelino-Camelia:2013gna,Amelino-Camelia:2013tla,Amelino-Camelia:2015dqa}, there appears to be an intimate connection between dimensional reduction to two, a non trivial geometry of energy-momentum space, scale invariance of quantum vacuum fluctuations and its relevance for the observed spectrum of fluctuations in the cosmic microwave background radiation. One of the priorities for future work will be a detailed analysis of these connections in the model presented here. 

Finally it should be noticed that, very recently, phase spaces with a compact momentum space have been shown to arise in loop quantum gravity when a positive cosmological constant is taken into account \cite{Rovelli:2015fwa}. While these results focus on the example of three spacetime dimensions, their relevance for the four-dimensional case is certainly interesting for the potential link between the model we presented here and quantum gravity scenarios.

\section*{Acknowledgements}
We would like to thank Jerzy Kowalski-Glikman for discussions. FN acknowledges support from CONACYT grant No. 250298.



\begin{thebibliography}{99}

\bibitem{Ambjorn:2005db}
  J.~Ambjorn, J.~Jurkiewicz and R.~Loll,
  Phys.\ Rev.\ Lett.\ {\bf 95}, 171301 (2005)
  [hep-th/0505113];
    O.~Lauscher, M.~Reuter,
  JHEP {\bf 0510}, 050 (2005)
  [hep-th/0508202].

\bibitem{Horava:2009if}
  P.~Horava,
  Phys.\ Rev.\ Lett.\ {\bf 102}, 161301 (2009)
  [arXiv:0902.3657 [hep-th]].

\bibitem{Carlip:2009kf} 
  S.~Carlip,
  AIP Conf.\ Proc.\  {\bf 1196}, 72 (2009)
  doi:10.1063/1.3284402
  [arXiv:0909.3329 [gr-qc]].
  
\bibitem{Modesto:2008jz} 
  L.~Modesto,
  Class.\ Quant.\ Grav.\  {\bf 26}, 242002 (2009)
  [arXiv:0812.2214 [gr-qc]].  

\bibitem{Benedetti:2008gu} 
  D.~Benedetti,
  Phys.\ Rev.\ Lett.\  {\bf 102}, 111303 (2009)
  [arXiv:0811.1396 [hep-th]].

\bibitem{Benedetti:2009ge}
  D.~Benedetti, J.~Henson,
  Phys.\ Rev.\ {\bf D80}, 124036 (2009)
  [arXiv:0911.0401 [hep-th]].

\bibitem{Sotiriou:2011aa}
  T.~P.~Sotiriou, M.~Visser and S.~Weinfurtner,
  Phys.\ Rev.\ D {\bf 84}, 104018 (2011)
  [arXiv:1105.6098 [hep-th]].

\bibitem{Calcagni:2013vsa}
  G.~Calcagni, A.~Eichhorn and F.~Saueressig,
  Phys.\ Rev.\ D {\bf 87}, 124028 (2013)
  [arXiv:1304.7247 [hep-th]].

\bibitem{Calcagni:2010bj}
  G.~Calcagni,
  JHEP {\bf 1003}, 120 (2010)
  [arXiv:1001.0571 [hep-th]].
  
\bibitem{Arzano:2013rka} 
  M.~Arzano and G.~Calcagni,
  Phys.\ Rev.\ D {\bf 88}, 084017 (2013)
  [arXiv:1307.6122 [hep-th]].  

\bibitem{Alesci:2012}
  E.~Alesci, M.~Arzano,
  Phys.\ Lett.\ B {\bf 707}, 272 (2012)
  [arXiv:1108.1507 [gr-qc]].

\bibitem{Arzano:2014jfa} 
  M.~Arzano and T.~Trzesniewski,
  Phys.\ Rev.\ D {\bf 89}, no. 12, 124024 (2014)
  [arXiv:1404.4762 [hep-th]].

\bibitem{Carlip:2015mra} 
  S.~Carlip,
  Class.\ Quant.\ Grav.\  {\bf 32}, no. 23, 232001 (2015)
  [arXiv:1506.08775 [gr-qc]].
  
\bibitem{Sotiriou:2010wn} 
  T.~P.~Sotiriou,
  J.\ Phys.\ Conf.\ Ser.\  {\bf 283}, 012034 (2011)
  [arXiv:1010.3218 [hep-th]].  
  
\bibitem{AmelinoCamelia:2000mn}
  G.~Amelino-Camelia,
  Int.\ J.\ Mod.\ Phys.\ D {\bf 11}, 35 (2002)
  [gr-qc/0012051]; 
  G.~Amelino-Camelia,
  Phys.\ Lett.\ B {\bf 510}, 255 (2001)
  [hep-th/0012238].
 
\bibitem{AmelinoCamelia:2003xp} 
  G.~Amelino-Camelia, L.~Smolin and A.~Starodubtsev,
  Class.\ Quant.\ Grav.\  {\bf 21}, 3095 (2004)
  [hep-th/0306134]. 

\bibitem{KowalskiGlikman:2004qa} 
  J.~Kowalski-Glikman,
  Lect.\ Notes Phys.\  {\bf 669}, 131 (2005)
  [hep-th/0405273].

\bibitem{Lukierski:1991}
  J.~Lukierski, H.~Ruegg, A.~Nowicki and V.~N.~Tolstoi,
  Phys.\ Lett.\ B {\bf 264}, 331 (1991).  
  
  \bibitem{Lukierski:1992}
  J.~Lukierski, A.~Nowicki and H.~Ruegg,
  Phys.\ Lett.\ B {\bf 293}, 344 (1992).

\bibitem{Lukierski:1994}
  J.~Lukierski, H.~Ruegg and A.~Nowicki,
  J.\ Math.\ Phys.\ {\bf 35}, 2607 (1994).

  \bibitem{Majid:1994}
  S.~Majid, H.~Ruegg,
  Phys.\ Lett.\ B {\bf 334}, 348 (1994)
  [hep-th/9405107].

\bibitem{Nowak:2002}
  J.~Kowalski-Glikman, S.~Nowak,
  Phys.\ Lett.\ B {\bf 539}, 126 (2002)
  [hep-th/0203040].

\bibitem{Kowalski:2002}
  J.~Kowalski-Glikman,
  Phys.\ Lett.\ B {\bf 547}, 291 (2002)
  [hep-th/0207279].

\bibitem{Kowalski:2003}
  J.~Kowalski-Glikman, S.~Nowak,
  Class.\ Quant.\ Grav.\ {\bf 20}, 4799 (2003)
  [hep-th/0304101].

\bibitem{KowalskiGlikman:2004tz} 
  J.~Kowalski-Glikman and S.~Nowak,
  hep-th/0411154.

\bibitem{Magueijo:2001cr} 
  J.~Magueijo and L.~Smolin,
  Phys.\ Rev.\ Lett.\  {\bf 88}, 190403 (2002)
  [hep-th/0112090].
  
\bibitem{Amelino-Camelia:2013cfa} 
  G.~Amelino-Camelia, M.~Arzano, G.~Gubitosi and J.~Magueijo,
  Phys.\ Lett.\ B {\bf 736}, 317 (2014)
  [arXiv:1311.3135 [gr-qc]].
  
\bibitem{Arzano:2014jua} 
  M.~Arzano, G.~Gubitosi, J.~Magueijo and G.~Amelino-Camelia,
  Phys.\ Rev.\ D {\bf 92}, no. 2, 024028 (2015)
  doi:10.1103/PhysRevD.92.024028
  [arXiv:1412.2054 [gr-qc]].  
  
\bibitem{Amelino-Camelia:2013gna} 
  G.~Amelino-Camelia, M.~Arzano, G.~Gubitosi and J.~Magueijo,
  Phys.\ Rev.\ D {\bf 88}, no. 10, 103524 (2013)
  [arXiv:1309.3999 [gr-qc]].  

\bibitem{'tHooft:1996uc} 
  G.~'t Hooft,
  Class.\ Quant.\ Grav.\  {\bf 13}, 1023 (1996)
  [gr-qc/9601014].
  
\bibitem{Welling:1997qz} 
  M.~Welling,
  Class.\ Quant.\ Grav.\  {\bf 14}, 3313 (1997)
  [gr-qc/9703058].
  

\bibitem{Carlip:1989nz} 
  S.~Carlip,
  Nucl.\ Phys.\ B {\bf 324}, 106 (1989).  

\bibitem{Bais:1998yn} 
  F.~A.~Bais and N.~M.~Muller,
  Nucl.\ Phys.\ B {\bf 530}, 349 (1998)
  [hep-th/9804130].
  
\bibitem{Bais:2002ye} 
  F.~A.~Bais, N.~M.~Muller and B.~J.~Schroers,
  Nucl.\ Phys.\ B {\bf 640}, 3 (2002)
  [hep-th/0205021].

\bibitem{Matschull:1997du} 
  H.~J.~Matschull and M.~Welling,
  Class.\ Quant.\ Grav.\  {\bf 15}, 2981 (1998)
  [gr-qc/9708054].  

  
\bibitem{Baez:2006un} 
  J.~C.~Baez, D.~K.~Wise and A.~S.~Crans,
  Adv.\ Theor.\ Math.\ Phys.\  {\bf 11}, no. 5, 707 (2007)
  [gr-qc/0603085].  
  
\bibitem{Arzano:2014ppa} 
  M.~Arzano, D.~Latini and M.~Lotito,
  SIGMA {\bf 10}, 079 (2014)
  [arXiv:1403.3038 [gr-qc]].  
  
\bibitem{Arzano:2016yir} 
   M.~Arzano and F.~Nettel,
  Phys.\ Rev.\ D {\bf 94}, no. 8, 085004 (2016)
  [arXiv:1602.05788 [hep-th]].
    
\bibitem{Arzano:2010jw} 
  M.~Arzano,
  Phys.\ Rev.\ D {\bf 83}, 025025 (2011)
  [arXiv:1009.1097 [hep-th]].    
    
\bibitem{Arzano:2014cya} 
  M.~Arzano,
  Phys.\ Rev.\ D {\bf 90}, no. 2, 024016 (2014)
  [arXiv:1403.6457 [hep-th]].   
    
\bibitem{Amelino-Camelia:2013tla} 
  G.~Amelino-Camelia, M.~Arzano, G.~Gubitosi and J.~Magueijo,
  Phys.\ Rev.\ D {\bf 87}, no. 12, 123532 (2013)
  [arXiv:1305.3153 [gr-qc]].  
  
\bibitem{Amelino-Camelia:2015dqa} 
  G.~Amelino-Camelia, M.~Arzano, G.~Gubitosi and J.~Magueijo,
  Int.\ J.\ Mod.\ Phys.\ D {\bf 24}, no. 12, 1543002 (2015)
  [arXiv:1505.04649 [gr-qc]].  

\bibitem{Rovelli:2015fwa} 
  C.~Rovelli and F.~Vidotto,
  Phys.\ Rev.\ D {\bf 91}, no. 8, 084037 (2015)
  [arXiv:1502.00278 [gr-qc]].    




\end{thebibliography}
\end{document}